\documentclass{aps2010} 
\usepackage{graphicx}

\title{Improving the spatial resolution by effective subtraction technique at Irkutsk incoherent scatter radar: the theory and experiment.}

\author[org1]{\textbf{\emph{Berngardt O.I.}}}
\author[org1]{\textbf{\emph{Kushnarev D.S.}}}

\address[org1]{Institute of Solar-Terrestrial physics SB RAS, 664033, Irkutsk, Lermontova Str., 126a, P.O.Box 291, Russia. berng@iszf.irk.ru}

\begin{document}
\maketitleblock  

\begin{abstract}
We describe a sounding technique that allows us to improve spatial
resolution of Irkutsk Incoherent Scatter Radar without loosing spectral
resolution. The technique is based on transmitting of rectangle pulses
of different duration in various sounding runs and subtracting correlation 
matrixes. Theoretically and experimentally we have shown, that subtraction 
of the mean-square parameters of the scattered signal for different kinds 
of the sounding signal one from another allows us to solve the problem 
within the framework of quasi-static ionospheric parameters approximation. 
\end{abstract}
\section{Problem}

The problem of the spatial resolution improvement without loosing spectral resolution 
is one of the basic technical problems of the incoherent scatter (IS) technique (as well 
as other remote sensing techniques, based on weak single backscattering from random media).
In such techniques the basic parameters that provide information about media properties are 
spectral power, correlation function and correlation matrix of the scattered signal (see [1]). 
One can show that first two functions can be easily obtained from the third one. From scattering theory 
it is also known that correlation matrix of the received signal is an average of the irregularities 
correlation function over the altitudes. The integration over altitudes is made with the weight defined by 
sounding signal shape and processing technique, and statistical averaging $<...>$ corresponds to the 
averaging over the sounding runs, see [1]: 
\begin{equation}
<u(T)\overline{u}(T+\tau)>=\int<W(r-\frac{cT}{2},\tau)P(r)\sigma(r,\tau)>dr;\, W(r,\tau)=a(\frac{rc}{2})\overline{a}(\frac{rc}{2}+\tau)\label{eq:1}\end{equation}
In the literature the integration weight $W(r,\tau)$ is referred as 'ambiguity function'. 
The power profile value $P(r)$ takes into account the radiowaves propagation, antenna pattern and 
cross-section dependence on radar range, $\sigma(r,\tau)$ - correlation characteristics of the 
scatterers and $a(t)$ - sounding signal shape. 

So, the practical problem of the spatial resolution improvement (formulated as obtaining information 
about correlation characteristics of the scatterers $\sigma(r,\tau)$ with integration over the smallest 
distances range) mathematically leads to the problem of the sounding and processing techniques synthesis  
that gives the ambiguity function $W(r,\tau)$ with necessary characteristics. 

The problem is currently under investigation (see [2]). But, for now, there are two most useful techniques 
for spatial resolution improvement: with using static amplitude-modulated signals (see [3]) and with 
phase-modulated signals, that change theirs shape from one sounding to another (see [4]). 
Constructing of these techniques is valid in the approximation of quasistationary media. 
In this approximation one can neglect changes of the irregularities correlation characteristics during averaging time 
and averaging over the sounding runs becomes equivalent to the statistical averaging. The use of such techniques allows, 
from one side, to obtain high spatial resolution, defined by discrete of temporal variation of the sounding signal, and 
from the other side to obtain correlation characteristics of the irregularities at large delays (lags), defined by 
duration of the whole signal. 

The well known difficulty of these techniques is impossibility to obtain the correlation function at all the lags with 
the same high spatial resolution and signal-to-noise ratio. This is caused by the following peculiarities of these 
techniques: correlation function of the irregularities at zero lag can not be obtained with high spatial resolution; 
correlation function can not be obtained at all the real lags within pulse duration, but only at discrete number of lags; 
the number of the 'correct' correlation function lags and spatial resolution are usually connected: the higher spatial 
resolution - the more lags in correlation function you get. So one can not obtian a lot of lags, necessary for correct 
estimation of the correlation function, when one need to obtain average spatial resolution (about 1/4 - 1/5 of the 
whole pulse length). 

We present sounding technique that allows to improve spatial resolution of incoherent scatter technique without the 
problems described above.

\section{Solution}

From (\ref{eq:1}) one can see that independence of irregularities correlation function on sounding signal shape allows 
one to use any linear combination of the correlation matrixes obtained during experiment instead of simple summarizing 
over the sounding runs(i.e. averaging over the sounding runs). In this case the ambiguity function will be 
(with accuracy of the custom multiplier) the same linear combination of the ambiguity functions for different sounding runs, 
summarized over the sounding run number $j$ (the fact, for example, is used in summation rules of the alternating code 
technique, see [4]): 
\begin{equation}
\sum D_{j}u_{j}(T)\overline{u}_{j}(T+\tau)=\int<W(r-\frac{cT}{2},\tau)><P(r)\sigma(r,\tau)>dr; 
\label{eq:2}\end{equation}
\begin{equation}
<W(r,\tau)>=\sum D_{j}a_{j}(\frac{rc}{2})\overline{a}_{j}(\frac{rc}{2}+\tau) \\
\label{eq:3}\end{equation}
The ideal ambiguity function must allow to determine irregularities correlation function without distortion effects, caused 
by power profile $P(r)$. At Irkutsk IS radar power profile is very irregular due to both electron density profile and Faraday 
variations profile (see [5]). One can show that the ideal ambiguity function must have the shape: 
\begin{equation}
<W(r,\tau)>=A(r)B(\tau)\label{eq:4}\end{equation}
We have found that for obtaining the almost ideal ambiguity function we should use a simple rectangle pulse as a sounding signal, 
with constant amplitude and with two custom durations, the pulse duration should change from one sounding to another: 
$dom (a_{2n})=[0,T_{long}];\, dom (a_{2n+1})=[0,T_{short}]$; 
during processing we should subtract the correlation matrix for shorter pulse from the correlation matrix for longer pulse. 
In this case $D_{2n}=1;D_{2n+1}=-1$ in (\ref{eq:2}). 
During the subtraction some parts of the ambiguity functions are effectively compensate each other, so we call this technique 
'effective subtraction technique' (EST). 
It is important to note that (for accurate positioning the ambiguity function) the correlation matrix must be calculated in the form: 
\begin{equation}
R_{j}(T,\tau)=\left\{ \begin{array}{c}
u_{j}(T-\tau)\overline{u}_{j}(T)\, for\,\tau\geq0\\
u_{j}(T)\overline{u}_{j}(T+\tau)\, for\,\tau<0\end{array}\right\} \label{eq:5}\end{equation}
During the experiment, we should calculate correlation matrix (\ref{eq:5}), but separately for sounding by the longer pulse 
$R_{long,2n}(T,\tau)$, and by shorter pulse $R_{short,2n+1}(T,\tau)$. The correlation matrix $R_{EST}(T,\tau)$ that provides 
ambiguity function with the almost ideal shape (\ref{eq:4}) is calculated as combination of $R_{long,2n}(T,\tau)$ and $R_{short,2n+1}(T,\tau)$: 
\begin{equation}
<R_{EST}(T,\tau)>=\sum R_{long,2n}(T,\tau)-\sum R_{short,2n+1}(T,\tau)\label{eq:5a}\end{equation}
The delay $T$ corresponding to the radar range should be measured from the sounding pulse start. 
Ambiguity functions of the correlation matrix (\ref{eq:5}) for longer pulse, shorter pulse and the difference between them 
(\ref{eq:5a}) are shown at Fig.1. From the Fig.1 one can see that ambiguity function of correlation matrix (\ref{eq:5a}) 
is close to the ideal one (\ref{eq:4}). One can show that the size of ambiguity function over the range (i.e. spatial resolution) 
is proportional to the difference of the pulses duration. 
Also, one can see, that the size of the ambiguity function over the lags (i.e. lag resolution) is proportional to the half-sum 
of the pulses durations. 
\begin{figure}
\label{fig1}
\includegraphics[scale=0.17]{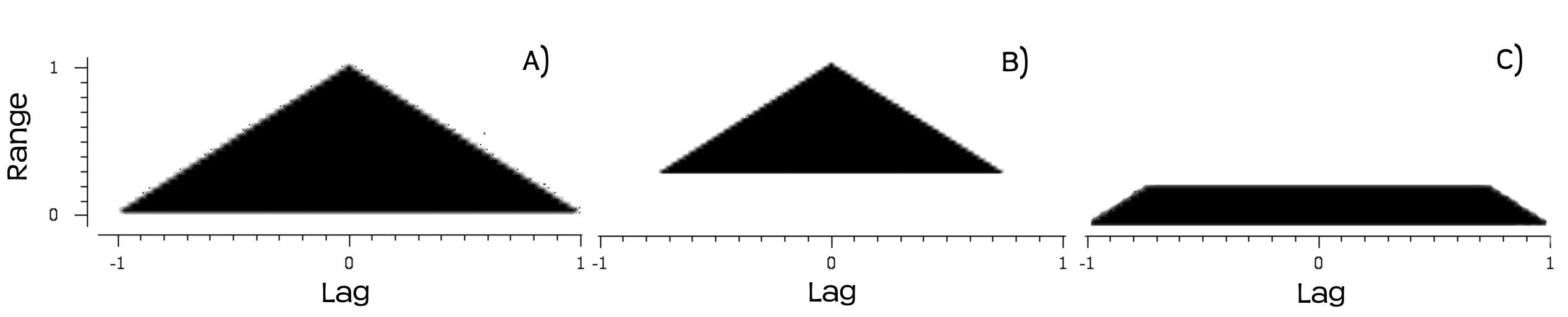}
\caption{Ambiguity function for EST (\ref{eq:5}-\ref{eq:5a}): for correlation matrix for longer pulse sounding (A),for 
correlation matrix for shorter pulse sounding(B), and for theirs difference (C). Range and lag are in longer pulse 
duration units.
}
\end{figure}

\section{Experimental results}

The EST was tested at Irkutsk IS radar [5] during the experiment 22-23/12/2010. As a sounding pulse we used a simple 
rectangle signal. Its duration was changed during sounding runs and was 900mks for longer pulse and 675mks for shorter pulse. 
The amplitude of the signal remains constant. Using received signal, we calculated correlation matrix (\ref{eq:5}) separately 
for longer pulse sounding runs $R_{long,2n}(T,\tau)$ and for shorter pulse $R_{short,2n+1}(T,\tau)$ sounding runs.
Correlation matrix that provides high spatial resolution was calculated as theirs combination (\ref{eq:5a}). 

For direct estimation of the ambiguity function we have used sounding runs with observations of the localized 
space scatterers. It can be shown that correlation matrix (\ref{eq:2}) in this case becomes close to the ambiguity function:
\begin{equation}
\sigma(r,\tau)=\delta(r-R_0); 
<R_{EST}(T,\tau)>=<P(R_0)><W(R_{0}-\frac{cT}{2},\tau)>
\end{equation}
The correlation matrix for localized scatterer (i.e. ambiguity function shape) obtained experimentally for EST is 
shown at Fig.2. As one can see, new technique actually improves the spatial resolution, as expected theoretically. 

\begin{figure}[t]
\label{fig2}
\begin{center}
\includegraphics[scale=0.54]{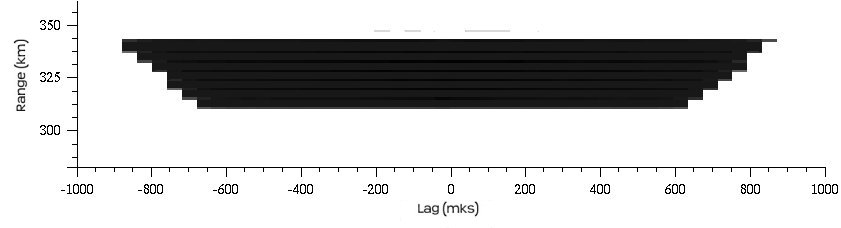}
\end{center}

\caption{The EST correlation matrix (\ref{eq:5}-\ref{eq:5a}) for scattering from localized space scatterer during experiment 23/12/2010.}
\end{figure}

At Fig.3 we present altitude dynamics of the spectral power for daytime and nighttime data, obtained with the technique 
described and with standard technique based only on longer pulse data. From the data presented one can see that spectral 
power valley near zero frequency for the new technique becomes deeper and can be observed even at the night data. 
It should be noted that such a spectral power shape actually theoretically expected for the ionospheric conditions 
(see [1]), so the new technique not only improves spatial resolution in comparison with standard technique, but slightly 
improves spectral resolution too.

\begin{figure}[t]
\label{fig3}
\begin{center}
\includegraphics[scale=0.23]{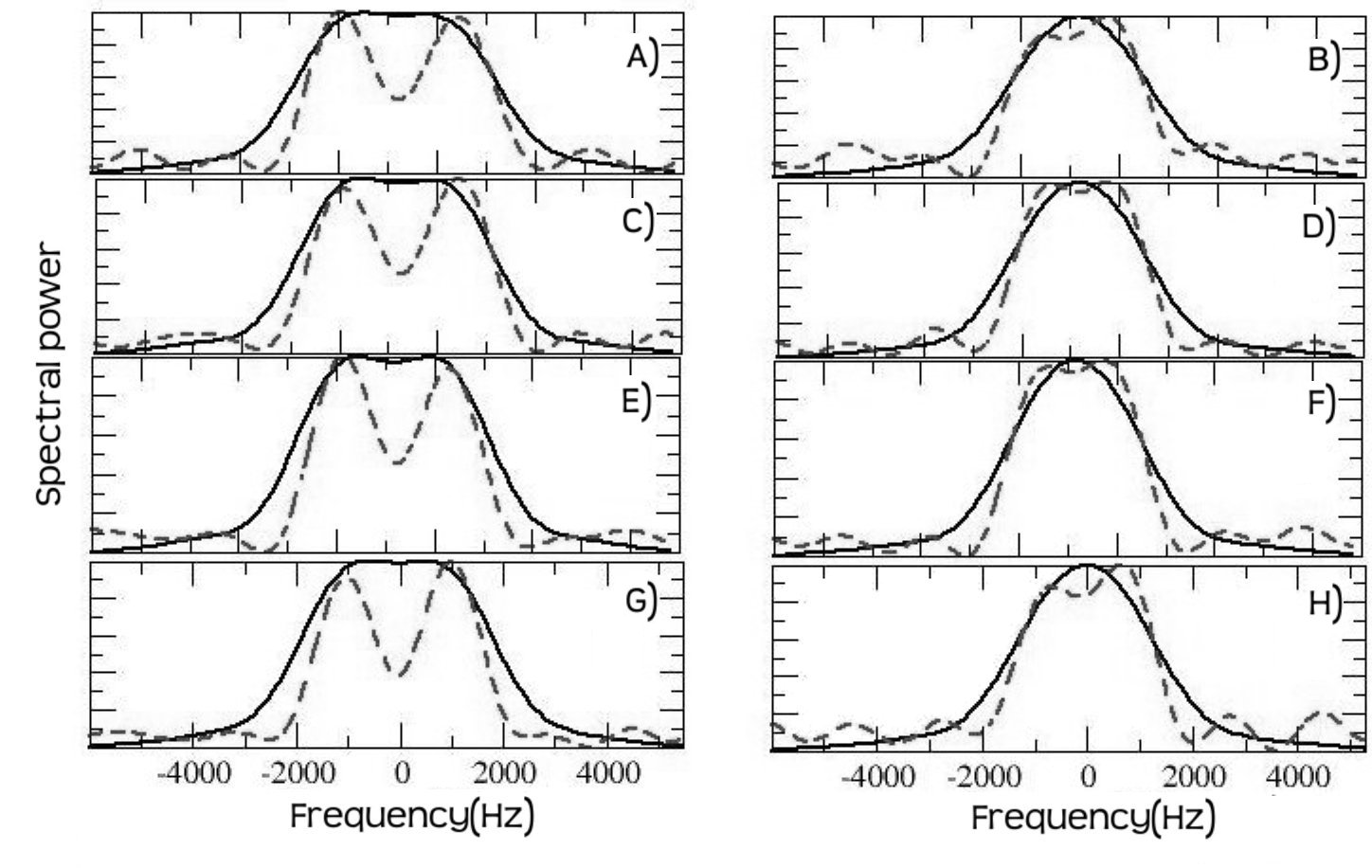}
\end{center}
\caption{Spectral power obtained with use of EST (dashed line), and standard Irkutsk IS radar technique (solid line) over the same data, 
obtained during experiment. (A,C,E,G) – daytime data, (B,D,F,H) – nighttime data. Altitudes: (G,H) - 230km, (E,F) - 260km, (C,D) - 290km, (A,B) - 320km}
\end{figure}

\section{Conclusion}

In conclusion we summarize some properties of the new effective subtraction technique (\ref{eq:5}-\ref{eq:5a}): 
the spatial resolution is equal for all the lags including zero lag, and correlation matrix can be calculated at 
any lag with almost equal signal-to-noise ratio; the EST uses only long enough simple pulses without any modulation, so in the 
case of the low signal-to-noise ratio (for example at high altitudes) we can use standard signal processing technique 
for both pulses separately with standard spatial resolution, and for high signal-to-noise ratio (at low altitudes) - 
use EST with high spatial resolution; the spatial resolution of EST is defined by pulse duration difference, and 
maximal allowed lags are defined by half-sum of the pulses durations.

\section{References}

1. Hagfors T. ``Plasma Fluctuations Excited by Charged Particle Motion and their 
Detection by Weak Scattering of Radio Waves" in D.Alcayde (ed.), 
\emph{Incoherent Scatter Theory, Practice and Science, Collection of lectures 
given in Cargese, Corsica, 1995. EISCAT Technical Report 97/53}, 1997, pp. 1-32.

2. M.S.Lehtinen , I.I.Virtanen, J.Vierinen, ``Fast comparison of IS radar code 
sequences for lag profile inversion" \emph{Ann.Geophysicae}, 26, 2008, 
pp. 2291-2301.

3. M.P.Sulzer, ``A radar technique for high range resolution incoherent scatter 
autocorrelation function measurements utilizing the full average power of 
klystron radars," \emph{Radio Sci}, 21(6), 1986, pp. 1033-1040.

4. A.Huuskonen, M.S.Lehtinen, J.Pirttil , ``Fractional lags in alternating codes: 
Improving incoherent scatter measurements by using lag estimates at noninteger 
multiples of baud length," \emph{Radio Sci.}, 32(6), 1996, pp. 2271-2282.

5. G.A.Zherebtsov, A.V.Zavorin, A.V.Medvedev , V.E.Nosov, A.P.Potekhin, 
B.G.Shpynev, ``Irkutsk Incoherent Scatter Radar," \emph{Radiotekh. Elektron. 
(Moscow)}, 47(11), 2002, pp. 1339–1345.

\end{document}